\documentclass{epl}

\usepackage{epsfig}
\usepackage{subfigure}
\title{
Free-Energy Barriers in the Sherrington-Kirkpatrick Model
}
\shorttitle{Free-Energy Barriers in the SK Model}
\author{Elmar Bittner and Wolfhard Janke}
\shortauthor{E. Bittner and W. Janke}

\institute{Institut f\"ur Theoretische Physik and Centre for
Theoretical Sciences (NTZ),\\
Universit\"at Leipzig, Augustusplatz 10/11, D-04109 Leipzig, Germany}


\begin{document}

\maketitle

\begin{abstract}
\noindent
The Sherrington-Kirkpatrick spin-glass model is investigated by means of
Monte Carlo simulations employing a combination of the multi-overlap 
algorithm with parallel
tempering methods. We investigate the finite-size scaling behaviour of the 
free-energy barriers which are visible in the probability density of the Parisi 
overlap parameter.  Assuming that the mean barrier height diverges with the 
number of spins $N$ as $N^\alpha$, our data show good agreement with the 
theoretical value $\alpha = 1/3$. 
\end{abstract}

%\section{Introduction}\label{intro}

Despite three decades of research the nature of the ``glassy''
low-temperature phase of finite-dimensional spin-glass systems 
remains a major open problem in statistical physics. It is
still unresolved whether the replica symmetry-breaking theory or the 
phenomenological droplet picture yields the correct description (for reviews,
see refs. \cite{young97,fischer,mezard,binder}). Even at the mean-field 
level, only very recently a mathematical proof~\cite{TalAca03} of Parisi's 
replica solution~\cite{parisi} for the Sherrington-Kirkpatrick (SK) model~\cite{sk} 
was given.

In the thermodynamic limit the frozen phase of the mean field spin glass shows 
many stable and metastable states. Such a feature is the consequence of the 
disorder and the frustration characterizing spin glasses in general and leading 
to a rugged free-energy landscape with probable regions (low free energy) 
separated by rare-event states (high free energy). But also for finite systems 
the free-energy landscape shows an intricate, corrugated structure. Therefore, 
it is hard to measure the free-energy barriers by means of conventional Monte 
Carlo simulations directly. The aim of this letter is to study the free-energy 
barriers of the SK mean field spin-glass model using the multi-overlap Monte 
Carlo algorithm~\cite{muq} and to compare our results with previous findings 
for finite-range spin glasses~\cite{bbabwj}. For analyzing the 
barriers, we used precisely the same method as introduced in
ref.~\cite{bbabwj} for the 
Edwards-Anderson (EA) nearest-neighbor model~\cite{ea}, 
where the scaling of the mean barrier height with the
number of spins was found to clearly deviate from
the theoretical mean-field prediction in both three and four
dimensions. To exclude the possibility that this deviation could, in principle, be caused by employing
different definitions in the theoretical and computational work, it is
important to apply precisely the same numerical procedure to the SK model.

%\section{Model and Simulation Techniques}\label{model}
% As we already mentioned we consider here the Sherrington-Kirkpatrick model with Hamiltonian
The Hamiltonian of the SK model reads
\begin{equation}  
H=-\sum_{i<j}J_{ij} s_i s_j~,
\end{equation}  
where $s_i=\pm1$, $i=1$,$\dots,N$, and the $J_{ij}$ are independent random variables
with a Gaussian distribution of zero mean and variance $N^{-1}$, $N$ being the 
numbers of spins. The critical temperature of the infinite system is $T_c=1$.

The fact that there is no parametrization of the relevant configurations by a
conventional thermodynamic variable led us to use the Parisi overlap parameter~\cite{parisi},
 \begin{equation}
q=\frac{1}{N}\sum_{i=1}^{N}s_i^{(1)} s_i^{(2)}~, 
 \end{equation}
as an order parameter, where the spin superscripts label two independent (real)
replicas for the same realization of randomly chosen exchange coupling constants
${\cal J}={J_{ij}}$. For given $\cal J$ the probability density of $q$
is denoted by $P_{\cal J}(q)$, and the function $P(q)$ is obtained as
\begin{equation}
P(q)=[P_{\cal J}(q)]_{\rm av}=\frac{1}{\#J}\sum_{\cal J}P_{\cal J}(q)~,
\end{equation} 
where $\#J$ is the number of realizations considered. For a given realization of
${\cal J}$ the nontrivial ({\em i.e.}, away from $q=\pm1$) minima are related to the
free-energy barriers of this disordered system ${\cal J}$. We are, therefore, 
interested in the whole range of the probability density $P_{\cal J}(q)$. 
Conventional, canonical Monte-Carlo simulations are not suited for such systems because
the likelihood to generate the corresponding rare-event configurations in the Gibbs
canonical ensemble is very small. This is overcome by non-Boltzmann 
sampling~\cite{berg_00,janke_98} with the multi-overlap weight~\cite{muq} 
\begin{equation}\label{w_q}
w_{\cal J}(q)=\exp\left[\beta\sum_{i<j} J_{ij} \left( s_i^{(1)} s_j^{(1)} +s_i^{(2)} 
s_j^{(2)} \right)
 + S_{\cal J}(q)\right]~,
\end{equation}
where the two replicas are coupled by $S_{\cal J}$
in such a way that a broad multi-overlap histogram $P_{\cal J}^{\rm muq}(q)$ over the entire
accessible range $-1 \le q \le 1$ is obtained.
When simulating with the multi-overlap weight, canonical expectation values of
any quantity ${\cal O}$ can be reconstructed by reweighting,
\begin{equation}
\langle {\cal O} \rangle^{\rm can}_{\cal J}=
\langle {\cal O} \exp(-S_{\cal J})\rangle_{\cal J}/\langle \exp(-S_{\cal J})\rangle_{\cal J}~.
\end{equation}
Ideally the weight function $W_{\cal J}\equiv\exp(S_{\cal J})$ should satisfy
\begin{equation}
P_{\cal J}^{\rm muq}(q)=P_{\cal J}^{\rm can} W_{\cal J}={\rm const.}~,
\end{equation}
{\em i.e.}, it should give rise to a completely flat multi-overlap probability density 
$P^{\rm muq}_{\cal J}(q)$. 
Of course, $P_{\cal J}(q)$ is {\em a priori} unknown and one has to proceed by iteration.
An efficient way to construct the weight function $W_{\cal J}$ is to use an accumulative recursion,
in which the new weight factor is computed
from all available data accumulated so far, for details see~refs.~\cite{berg96} and ~\cite{wj_nic}.
The multi-overlap algorithm combined with this recursion allows an almost automatic
simulation of the SK model.

Let us now turn to the description of the Monte Carlo update procedure used
by us. We combined the multi-overlap algorithm~\cite{muq} as described above 
with the parallel tempering optimized Monte Carlo procedure~\cite{pt} 
to overcome as many as possible hidden barriers in the rugged phase space.
We studied systems with $N=32, 64, 128, 256, 512$ and $1024$ spins. For the parallel tempering 
procedure we chose a set of $32$ temperature values in the range $T=1/3 - 1.6$ for all of
our systems apart from the largest, where we used $64$ temperature values
for the same temperature interval. 
Once the entire range of $q$ for all temperature values
was covered, the accumulative recursion for the weight functions was stopped. 
Due to large differences in the free-energy landscape for different disorder 
realizations ${\cal J}$, the number of recursion steps varied for different ${\cal J}$.
After the weight functions were constructed, they were kept fixed and we took about
$100\,000$ measurements, with five sweeps between the measurements. A sweep consisted of
$N$ spin flips with the multi-overlap algorithm and one parallel tempering update.
We recorded time series of the overlap parameter $q$ for five different temperature values
and the canonical $P_{\cal J}(q)$ distribution for all temperature values.
To average over the disorder we used
$1000$ realizations of the disorder for $N\le512$ and $100$ for $N=1024$. 

For each of these samples ${\cal J}$ we computed the barrier autocorrelation time $\tau_B$
by employing
the same method as Berg {\it et al.} used for the $\pm J$ EA Ising 
spin-glass model~\cite{bbabwj}.
For clarity, we recall the basic idea here.
The free-energy barrier $F_B$ for a given $P_{\cal J}(q)$
is defined through the autocorrelation time of a one-dimensional Markov process which
has the canonical $P_{\cal J}(q)$ distribution as equilibrium state.
The transition probabilities $T_{i,j}$ are given by
\begin{equation} \label{T}
T=\left[\begin{array}{cccc}1-w_{2,1}&w_{1,2}&0 & \ldots\cr
            w_{2,1}&1-w_{1,2}-w_{3,2}&w_{2,3}&\ldots\cr
            0&w_{3,2}&1-w_{2,3}-w_{4,3} &\ldots\cr
            0 & 0 & w_{4,3} & \ldots\cr
            \vdots&\vdots&\vdots&\ddots\cr\end{array}\right]~,
\end{equation}
where $w_{i,j}$ ($ i \neq j)$ is a   probability {\em \`a la} Metropolis 
to jump from state $q=q_j$ to $q=q_i$ ($q_i=i/N$, 
$i\in [-N , -N+2 , \dots , +N]$),
\begin{equation} 
 w_{i,j}={1\over 2} 
 \min\Bigl(1,{P_{\cal J}(q_i)\over P_{\cal J}(q_j)}\Bigr)~.
\end{equation}  
The transition matrix $T$  fulfils the detailed balance condition (with $P_{\cal J}$), and
as a consequence it has only real eigenvalues. The largest eigenvalue 
(equal to one) is non-degenerate, and the second largest eigenvalue 
$\lambda_1$ determines the autocorrelation time of the Markov chain,
\begin{equation} \label{tauq}
\tau_{B,{\cal J}} = {1\over{N (1-\lambda_1)}}~.
\end{equation} 
The  associated free-energy barrier for realization
${\cal J}$ is defined as
\begin{equation} \label{F_B^q}
F_{B,{\cal J}}=\ln(\tau_{B,{\cal J}})~. 
\end{equation}
Note that the definition of the autocorrelation time~(\ref{tauq}) takes only barriers in $q$ into account,
but not other barriers which may well exist in the multidimensional configuration space.

\begin{figure}[t]
\vspace{-4mm}
\centerline{\psfig{figure=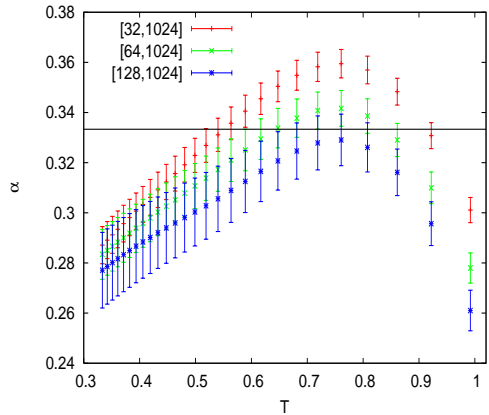,angle=0,height=5.66cm,width=6.82cm}}
\caption{\label{fit}
Dependence of the exponent $\alpha$ in the ansatz $F_B=c N^\alpha$ on the lower bound of the fit 
range $[N_{\rm min},1024]$ as a function of temperature. The horizontal line
indicates the theoretical value $\alpha=1/3$. 
}
\end{figure}

For each temperature value we performed least-squares fits of the finite-size scaling (FSS)
ansatz $F_B\equiv[F_ {B,{\cal J}}]_{\rm av}=c N^\alpha$ which corresponds to the exponential FSS behaviour
\begin{equation}
\tau_B\propto e^{c N^\alpha}~.
\end{equation}
The results from these fits are depicted in fig.~\ref{fit} and are 
consistent with previous results in the 
literature~\cite{young82,nemoto,rodgers,vertechi,colborne,kinzelbach,takayama,billoire}
using numerical and analytical methods. The horizontal line in fig.~\ref{fit} indicates
the theoretical value $\alpha=1/3$ of ref.~\cite{kinzelbach}. The three data sets
show fits with different lower bounds of the fit range $N_{\rm min}$, 
while the upper bound was always our largest system $N=1024$. From these fits we see
a strong finite-size effect for $T \to T_c=1$.
At lower temperatures we find a linearly increasing deviation from the theoretical value. 
This is also a finite-size effect, because the slope of the deviation becomes flatter 
when increasing the lower bound of the fit range and there is no physical reason for a
change of behaviour of the barrier autocorrelation time in the glassy phase.

\begin{figure}[t]
\vspace{-4mm}
\centerline{\psfig{figure=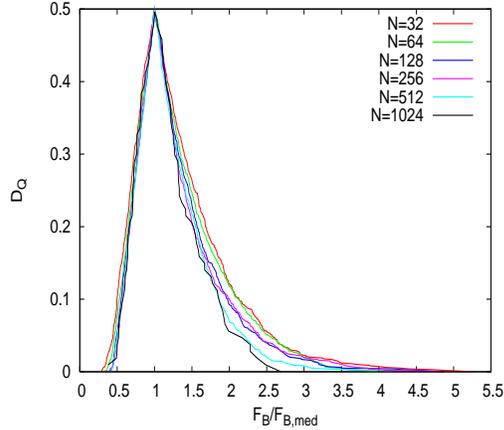,angle=0,height=5.85cm,width=6.82cm}}
\caption{\label{non_self}
Probability distribution function $D_Q$ for the free-energy barriers at $T=1/3$ in units 
of their median value. 
}
\end{figure}

One possible explanation for this deviation from the theoretical value is the lack of
self-averaging of the finite volume Parisi overlap parameter distribution 
$P_{\cal J}$ in the SK model~\cite{pastur}.
To check the non-self-averaging of our numerical data, we analyzed our free-energy barriers 
relying on the (empirical) cumulative distribution function $D(x)$,
which is defined for a set of sorted data, e.g., the free-energy barriers
$F_{B,1}<F_{B,2}<\dots<F_{B,n}$, by 
\begin{equation} 
\frac{i}{n}-\frac{1}{2n}\le D(x)\le \frac{i}{n}+\frac{1}{2n} \quad {\rm for} \quad F_{B,i}\le x \le F_{B,i+1}~,
\end{equation} 
where we use a straight-line interpolation in between. A nice way to test the non-self-averaging 
is to look at the peaked distribution function~\cite{bbabwj}:
\begin{equation}
 D_Q (x) = \cases{ D(x)\quad\quad  {\rm  for}\ D(x)\le 0.5~,\cr
               1 - D(x)\  {\rm  for}\ D(x)\ge 0.5~.\cr}
\end{equation}
This function has a peak at the median $x_{\rm med}$ of the data with $D_Q=0.5$.
For self-averaging data $x$, the function $D_Q$ collapses in infinite volume to
%$$ D_Q (x) = \cases{ 0.5\ {\rm for}\ x=\overline{x}\cr
%                     0\ {\rm otherwise}~,\cr} $$
$D_Q (x) = 0.5$  for $x=\overline{x}$ and 0 otherwise,
where $\overline{x}$ is the mean value. For non-averaging 
quantities the width of $D_Q$ stays finite. In fig.~\ref{non_self} we show the
peaked distribution function in units of the median value which also scales approximately 
with $N^{1/3}$.
We observe a very weak finite-size dependence %of the peaked distribution function
and therefore clearly a non-self-averaging behaviour of the free-energy barriers. 
This result is in agreement with theoretical calculations~\cite{pastur}.
The width of $D_Q$ increases with decreasing temperature, {\em i.e.} the sample to sample variations
become more pronounced for lower temperatures.

\begin{figure}[b]
\vspace{-4mm}
\centerline{\psfig{figure=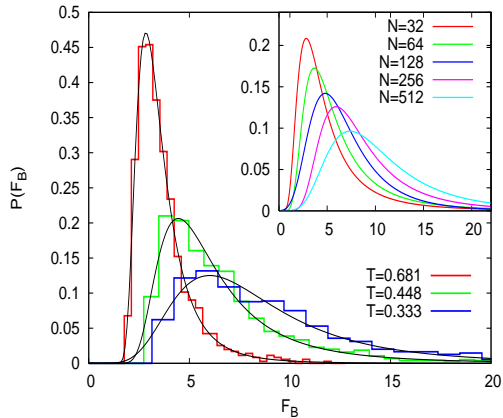,angle=0,height=5.66cm,width=6.82cm}}
\caption{\label{frechetp1}
Density of free-energy barriers $F_B$ for $N=256$ at different temperatures. 
The inset shows the densities for $T=1/3$ for different numbers of spins.
}
\end{figure}

\begin{figure}[t]
\vspace{-3mm}
\centerline{\psfig{figure=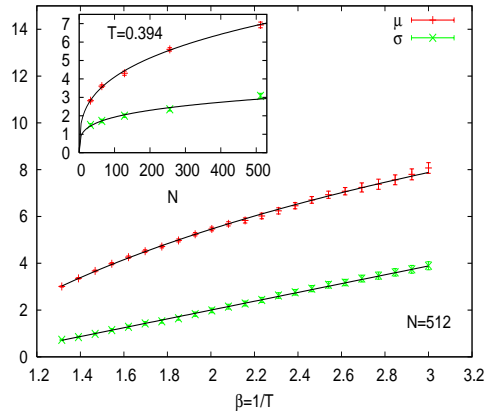,angle=0,height=5.66cm,width=6.82cm}}
\caption{\label{frechetp2} Temperature dependence of the parameters $\mu$ and $\sigma$ of the
Fr\'echet distribution for $N=512$. The inset shows the size dependence of $\mu$ and $\sigma$
for $T=0.394$.}
\end{figure}

We already mentioned that the distribution of the free-energy barriers becomes broader for
low temperatures. Now let us have a closer look at the distribution itself, therefore we only 
analyze the system sizes where we have $1000$ different disorder realizations. 
In recent work, Dayal~{\it et al.}~\cite{dayal} have found that the tunnelling times of their 
flat-histogram sampling simulations of the 2D $\pm J$ EA Ising spin glass are
distributed according to the Fr\'echet extremal value distribution for fat-tailed 
distributions.
In general, extreme-value statistics can be classified into different universality 
classes~\cite{gumbel,galambos}, depending on whether the tails of
the original distribution are fat tailed (algebraic),
exponential, or thin tailed (decaying faster then exponential). 
Assuming that the free-energy barriers of the SK model are, as well as
the tunnelling times of the EA spin glass, distributed according to an extreme-value distribution,
we use the integrated probability density of the generalized extreme value distribution (GEV),
\begin{equation}\label{frechet}
F_{\xi;\mu;\sigma}(x)=\exp\left[-\left(1+\xi\frac{x-\mu}{\sigma}\right)^{-1/\xi}\right] 
\end{equation}
for~$1+\xi(x-\mu)/\sigma>0$,~to fit our data. Here,~$\xi$~is the shape parameter, and~$\mu$~and~$\sigma$
are related to the $N$-dependent mean and variance of the distribution,~respectively.~We
find that the free-energy barriers show fat
tails for $T<T_c$ with shape parameter $\xi>0$, {\em i.e.}, a Fr\'echet distribution. In fig.~\ref{frechetp1}
we plot the resulting fits and data of the probability density for different temperatures
below the glass transition and  
find that the tails become fatter and fatter as the temperature goes to zero. 
The histograms for low temperatures show deviations from the Fr\'echet distribution
for small values of $F_B$, so a much larger number of disorder realizations would be
needed to determine both tails of the distribution properly. 
We determined the parameters $\mu$, $\sigma$ and $\xi$ for different temperatures and found that
$\mu$ grows logarithmically and $\sigma$  linearly with inverse temperature $1/T$, whereas
$\xi$ stays more or less constant at $\xi\approx 0.33$. As an example we show in fig.~\ref{frechetp2} the results
for $N=512$.
If we keep the temperature fixed and look at the size dependence of
the distribution, we find that for a larger number of spins the distribution becomes broader,
c.f.\ the inset of fig.~\ref{frechetp1}. To quantify this behaviour we use the scaling 
relations $\mu \propto N^{\alpha_\mu}$ and $\sigma \propto N^{\alpha_\sigma}$, 
which lead to $\alpha_\mu\approx 0.31$ and $\alpha_\sigma\approx0.25$ for our lowest 
temperatures, see the inset of fig.~\ref{frechetp2}. 
The exponent of the location parameter $\mu$ is closely related to the scaling 
exponent of the mean barrier height, therefore
$\alpha_\mu$ should be $1/3$, which is in good agreement with our result.
We find a weak temperature dependence of the exponents
$\alpha_\mu$ and $\alpha_\sigma$ with positive and negative slope for increasing $T$, respectively.

As a short remark we want to mention that we also looked at the correlation between the $n$
largest eigenvalues $\lambda_n$ of the interaction matrix $J_{ij}$, which is a real symmetric matrix, and the 
largest free-energy barrier $F_B$ for this disorder realisation. And we find for $T < T_c$ a very weak but non-vanishing correlation between the largest 
eigenvalue $\lambda_1$ and $F_B$. Unfortunately,
this correlation becomes weaker and weaker for larger systems and also for smaller temperatures.
All other correlations vanish much faster, such that these correlations cannot be used to predict
the magnitude of the free-energy barrier based only on $J_{ij}$.

To conclude, we found that the free-energy barriers of the SK model are non-self-averaging
and distributed according to the Fr\'echet extremal value distribution. 
These particular features were also found for the EA nearest-neighbor model and such 
similarities support the position that the Parisi replica symmetry breaking solution 
of the SK model is the limit of the short-range model on a lattice in dimension $d$ when
$d \to \infty$, with a proper rescaling of the strength of the Hamiltonian.  
On the other hand,
we also found that the free-energy barriers diverge with the theoretically predicted 
value $\alpha=1/3$, which is in contrast to the results for the EA model in three and four 
dimensions~\cite{bbabwj}. Of course, one reason for this discrepancy could possibly be
that finite-size effects, caused by the relative large temperature used in ref.~\cite{bbabwj},
are too strong or the lattice sizes are too small to see the real asymptotic behaviour. Still, 
our results also support the complementary position, that 
the Parisi replica symmetry breaking solution sheds little light on the thermodynamic structure of
the EA model~\cite{newman}. With these oppositional results we are not able to rule out
one of the two scenarios. But we can state that the method to determine the free-energy barriers
proposed by Berg {\it et al.}~\cite{bbabwj} leads to correct results for the SK model 
and is very useful for systems with many barriers in the free energy.

%%%%%%%%%%%%%%%%%%%%%%%%%%%%%%%%%%%%%%%%%%%%%%%%%%%%%%%%%%%%%%%%%%%%%%%%%%
 \acknowledgments
We gratefully acknowledge financial support from the Deutsche Forschungsgemeinschaft (DFG)
under Grant No. JA 483/22-1 and the EU RTN-Network `ENRAGE': {\em Random Geometry
and Random Matrices: From Quantum Gravity to Econophysics\/} under grant
No.~MRTN-CT-2004-005616.
%%%%%%%%%%%%%%%%%%%%%%%%%%%%%%%%%%%%%%%%%%%%%%%%%%%%%%%%%%%%%%%%%%%%%%%%%%

\end{document}